\documentclass[preprint,
superscriptaddress,
 amsmath,amssymb,
 aps,
pra,
]{revtex4-2}

\usepackage{graphicx}
\usepackage{bm}
\usepackage{xcolor}
\usepackage[separate-uncertainty=true, separate-uncertainty-units=single]{siunitx}

\begin{document}

\title{The Coming of Age of the Ultracold Electron Source: A Review}

\author{Julius Huijts}
    \email{j.v.huijts@tue.nl}
    \affiliation{Department of Applied Physics and Science Education\\
    Eindhoven University of Technology, P.O. Box 513, 5600 MB Eindhoven, The Netherlands}

\author{Jom Luiten}
    \affiliation{Department of Applied Physics and Science Education\\
    Eindhoven University of Technology, P.O. Box 513, 5600 MB Eindhoven, The Netherlands}

\date{\today}

\begin{abstract}
The ultracold electron source is a unique approach to the generation of high-brightness electron beams. We give an overview of its development over the past 20 years, including the underlying physical principles, technical details and recent experiments, and give a flavor of the exciting prospects that the future may hold.
\end{abstract}

\maketitle

\section{Introduction}
Two decades ago, a radically different approach to the generation of high-brightness, low-emittance electron beams was born: based on the photo-ionization of laser-cooled atoms \cite{claessens2005ultracold}. Dubbed the UltraCold Electron Source or 'UCES', it is characterized by an extremely low (order \qty{10}{K}) transverse electron temperature, promising for the generation of high-brightness electron beams. Such beams find applications in e.g. ultrafast electron diffraction (UED) \cite{siwick2003atomic,vanoudheusden2010compression,filippetto2022ultrafast}, compact X-ray sources based on inverse Compton scattering \cite{graves2014compact,vanelk2025xrayscompacttunablelinacbased}, or even as injectors for free-electron lasers (FEL) \cite{akre2008commissioning,vanderGeer2014ultracold}.
Now, twenty years later, the time is ripe to give a complete overview of how the UCES-technology developed, and what promises the future may still hold. While a previous review (\cite{mcculloch2016review}) gave an overview of the field in 2016, the present work provides much greater technical detail and includes the developments of the past decade.

We restrict this review to the generation of electron bunches through photoionization of neutral atoms in a magneto-optical trap (MOT). The same principle can be used to produce ultracold ion bunches - this is treated elsewhere (\cite{hanssen2006laser,mcclelland2016bright,tenhaaf2018measurements}). For the closely related technique of creating electron beams through (photo- or field) ionization of neutral atom beams, see e.g.
\cite{gallagher1974photoionization, kurokawa2010threshold, mcculloch2017field}. 

The paper is organized as follows. In Section \ref{sec:bunchquality} we first recall the theory on electron bunch quality and show why it is attractive to go to low beam temperatures. Then we describe the ``Birth" of the ultracold electron source in Section \ref{sec:birth}: the idea, the principle and its first realizations. In section \ref{sec:youth} we will see how the application of femtosecond pulses in the ionization process lead to lower bunch temperatures than initially expected, sparking a series of studies to achieve a proper understanding of the electron emission process. Dubbed ``Youth: play and learn", this section also covers performed experiments on phase-space shaping and the first electron diffraction experiments. Section \ref{sec:adolescence} covers the more recent developments towards usable instruments, thus titled ``Adolescence". We end with Conclusions (Section \ref{sec:conclusion}) and an overview of exciting future research directions (Section \ref{sec:outlook}).

\section{Electron bunch quality}\label{sec:bunchquality}
\subsection{Brightness and emittance}
The ideal electron bunch carries a large amount of charge with a narrow energy spread, in a short time, and can be focused down to a tiny spot. The metric that quantifies this is the bunch brightness:

\begin{equation}\label{eq:bunchbrightness}
    \hat{B}_n \equiv \frac{1}{mc} \frac{N e}{(2\pi)^3 \epsilon_x\epsilon_y\epsilon_z},
\end{equation}
where $N$ is the number of electrons in a bunch, $e$ and $m$ the electron charge and mass respectively. $\epsilon_{x,y,z}$ are the emittances in the corresponding directions, and represent the phase-space volume occupied by the bunch. Here, the longitudinal emittance (typically $\epsilon_z$) roughly represents the product of bunch length and energy spread, while the transverse emittances ($\epsilon_x$ and $\epsilon_y$) correspond to the focusability of the beam, analogous to the \textit{étendue} or beam parameter product (BPP) in light optics.

More formally, the (normalized) emittance is defined as
\begin{equation}\label{eq:emittance}
    \epsilon_x \equiv \frac{1}{mc}\sqrt{\left< x^2 \right> \left< p_x^2 \right> - \left< xp_x \right>^2},
\end{equation}
where $x$ and $p_x$ are the position and momentum of each electron (and equivalently for $y$ and $z$), and the brackets denote averaging over the entire bunch. 
In other words, the emittance is the determinant of the covariance matrix of the beam's phase space coordinates, and thus a measure of the volume that the beam occupies in this phase space. Following Liouville's theorem, the emittance is conserved as the electron bunch propagates through a system, although in practice it will increase monotonically.
In absence of correlations between position and momentum - as is the case for the transverse dimensions at the source or at a focus - the expression for the emittance reduces to the simple product between beam size and momentum spread:
\begin{align}\label{eq:emittance_nocorrelation}
    \epsilon_x  &=  \frac{1}{mc}\sqrt{\left< x^2 \right> \left< p_x^2 \right>}\quad\text{(at source or focus)} \nonumber\\
    &= \frac{\sigma_x\sigma_{p_x}}{mc}.
\end{align}
It is instructive to rewrite the longitudinal emittance in units of time and energy using $\sigma_z\approx c\sigma_t$ and $\sigma_{p_z}/mc=\sigma_U/mc^2$, such that:
\begin{equation}\label{eq:eps_z}
    \epsilon_z=\frac{\sigma_t\sigma_U}{mc}
\end{equation}
 with $\sigma_U$ the rms energy spread of the beam.
Often (for many cases in UED for example) this energy spread is deemed less important, and the 5-D or transverse brightness is used instead:
\begin{align}\label{eq:transversebrightness}
    B_\perp &\equiv \frac{\hat{I}}{(2\pi)^2\epsilon_x\epsilon_y} \nonumber \\
    &= \frac{Q}{(2\pi)^{3/2}\epsilon_x\epsilon_y\sigma_t}. \quad \text{(assuming Gaussian temporal profile)}
\end{align}

\subsection{$\sigma_{p_x}$: Temperature, MTE and coherence length}
The momentum spread $\sigma_{p_x}$ appears in different forms in different domains of science. From a statistical physics point of view, $\sigma_{p_x}$ describes the random motion of the particles, and can therefore be intuitively expressed as a beam temperature. Indeed, for a thermal, Gaussian distribution:
\begin{equation}
    \sigma_{p_x}^2=\sigma_{p_y}^2=\sigma_{p_z}^2=mk_BT.
\end{equation}
This means that the emittance can also be expressed as
\begin{equation}\label{eq:emittance_T}
    \epsilon_x = \sigma_x \sqrt{\frac{k_BT}{mc^2}}.
\end{equation}
In photocathode development, the transverse momentum spread is commonly expressed as an energy, known as the Mean Transverse Energy:
\begin{equation}
    MTE = \frac{\left < p_x^2 \right >}{2m} = \frac{\sigma_{p_x}^2}{2m},
\end{equation}
which equals $k_BT/2$.
And finally, to those seeking to apply a beam to UED experiments for example, it is more intuitive to talk about the transverse coherence length:
\begin{align}
    L_\perp = \frac{\hbar}{\sigma_{p_x}},
\end{align}
which needs to be larger than the lattice constant such that constructive interference can occur and a diffraction pattern can be formed. It is intuitive to look at the ratio of the coherence length to the transverse beam size as a definition of the relative coherence of a beam:
\begin{align}
    C_\perp &= \frac{L_\perp}{\sigma_x},\\
    &=\frac{\hbar}{\epsilon_x mc} = \frac{\lambda_C}{\epsilon_x},\nonumber
\end{align}
which, as the second line shows, is inversely proportional to the emittance. In the last step the Compton wavelength $\lambda_C=\hbar/mc\approx$ \qty{0.4}{pm} is introduced.

\subsection{Routes towards high-brightness beams}
In the quest for brighter beams, the most obvious approach may seem to try to extract a high charge from a small source size, to maximize current density at the source. It is indeed a common approach with conventional photocathodes to minimize $\sigma_x$ in order to produce low-emittance, high-brightness beams, and techniques like laser-triggered emission from nanotips successfully take this approach to the extreme \cite{hommelhoff2006field,ropers2007localized}. However, this approach runs into two limitations. First, the extraction of a charge $Q$ from a surface gives rise to an image charge field of $E_{image} = Q/(2\pi\epsilon_0 \sigma^2_{x,source})$. This image charge field has to stay small compared to the accelerating field (by at least an order of magnitude\cite{vanOudheusden2007electronsource}). For typical accelerating fields of tens of MV/m that means $Q/\sigma^2_{source} < 100$ attoCoulomb/\unit{\um^2}. 

The second limitation is given by the Coulomb repulsion of the electrons, or space charge. These space charge fields are of a magnitude comparable to $E_{image}$ \cite{vanOudheusden-thesis}, but their effect on the emittance depends drastically on the profile of the bunch density. Only for uniformly filled ellipsoidal (or `waterbag') bunches are the space charge forces strictly linear in space, avoiding emittance growth \cite{kapchinskii1959proceedings,luiten2004realize}. All other density profiles lead to emittance growth, which can quickly become dominant even at moderate bunch densities. For example, `pancake'-shaped bunches common in UED, with a Gaussian transverse profile with typical values of $\sigma_{source}=100\,\unit{\um}$ and $\sigma_z=0.4\,\unit{\um}$, already show emittance degradation due to transverse shock formation at a charge of just 1000 electrons \cite{zerbe2018dynamical}, corresponding to a limit on $Q/\sigma^2_{source}$ that is 4 orders of magnitude lower than that posed by the image charge.

Instead of minimizing $\sigma_x$, the alternative approach towards the realization of a bright beam is to minimize $\sigma_{p_x}$ instead, i.e. to minimize the beam temperature at the source. In modern photocathode development, this is now known as "intrinsic emittance reduction".
For continuous electron beams, this principle was already applied as early as 1991, in the production of cold electron beam for the stochastic cooling of ion beams at MPIK (see \cite{zwickler1991photocathode}). But to the best of our knowledge, the UltraCold Electron Source was the first pulsed electron source to take this approach, and as this review will show, it takes this approach to the extreme.

\section{The birth of the ultracold electron source}\label{sec:birth}
The ultracold electron source was born out of the merger of two research groups at the TU Eindhoven: one focusing on accelerator physics and applications, and the other on atomic physics and quantum technology. Soon the idea rose to investigate the potential of using an ultracold plasma (UCP) \cite{killian1999creation} as a high-brightness source of electrons.

\subsection{The principle}\label{sec:principle}
In the past 20 years of development, the main principle behind the ultracold electron source has stayed the same. The UCES seminal paper \cite{claessens2005ultracold} described it as follows, with reference to figure \ref{fig:Claessensprinciple}:

\begin{figure}
    \centering
    \includegraphics[width=0.5\linewidth]{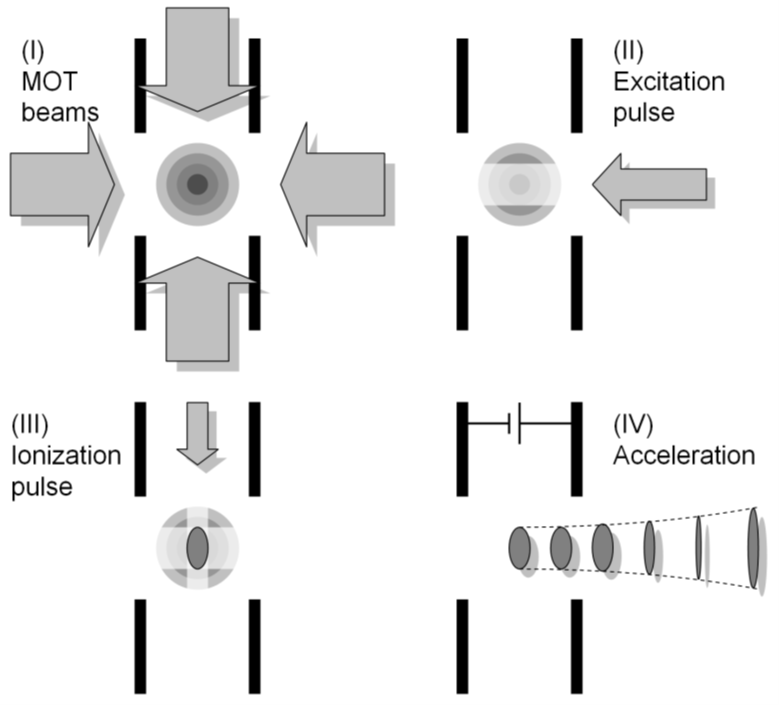}
    \caption{Schematic of the four-step procedure to realize a pulsed UCP electron source. From \cite{Claessens-thesis}.}
    \label{fig:Claessensprinciple}
\end{figure}

\begin{description}
    \item[(I)] A cold ($T < 1\,\text{mK}$) cloud of atoms is trapped in a magneto-optical trap (MOT) \cite{raab1987trapping} in a volume of a few mm$^3$.
    \item[(II)] Part of the cold atom cloud is excited to an intermediate state with a quasicontinuous \unit{\us} pulse.
    \item[(III)] Then, a pulsed-laser beam propagating at right angles to the excitation laser ionizes the excited atoms only within the volume irradiated by both lasers. Here a UCP is formed.
    \item[(IV)] The bunches are extracted by an electric field.
\end{description}

While this main principle has stayed the same, the experimental details on today's realization of this source differ considerably from the original proposition. Before describing that evolution we would first like to note four important characteristics unique to the UCES, that were already realized from its inception.

First, the fact that the electrons are created over a volume (instead of a surface like with conventional photocathodes) means that electrons in the back of the bunch experience a larger acceleration potential difference than the electrons in the front of the bunch. Thus, the accelerated electron bunches go through a self-compression point (as indicated in step \textbf{IV} in Fig. \ref{fig:Claessensprinciple}). The initial simulations already predicted sub-ps bunch lengths at this point, which was later confirmed in experiment \cite{deRaadt2023subpicosecond} as we will see in Section \ref{sec:subps}. If the bunches are accelerated over a distance $d_{acc}$, the self-compression point lies at $2d_{acc}$ after the accelerator. 

Second, as the electron bunch is created at the overlap between the excitation and ionization laser beams, the initial charge distribution can be controlled in 3D. This offers a direct way to create the uniformly filled ellipsoidal bunches mentioned above, suppressing emittance degradation caused by space charge. 

Third, while most solid state electron emitters require conditioning and suffer from aging, with the UCES each electron bunch is essentially generated from a `fresh' cathode.

Finally, it is actually not strictly necessary to use a UCP in order to achieve low electron temperatures \cite{luiten2007ultracold}. The initial transverse momentum spread (or temperature, or MTE) is dominated by the excess energy of the ionizing photons. Even if an atomic gas at room temperature was used, the electrons are extracted well before thermal equilibrium between the electrons and ions is achieved. The higher temperature of the atoms before photoionization does cause a Doppler broadening of the photoionization threshold, which for room temperature atoms would be on the order of 10 K. In other words: as long as the electron extraction is sufficiently fast, a 10 K electron beam can in principle also be produced by near-threshold photo-ionization of a room temperature atomic gas. ``Sufficiently fast" here means compared to the plasma period $T_p=2\pi\sqrt{\frac{\epsilon_0 m}{n_e e^2}}$, which is on the order of a nanosecond for our plasma densities. 
There are however two reasons to use a UCP, one practical and one fundamental. First, a MOT allows to confine a high density of neutral atoms in a high-vacuum environment, such that the accelerated electrons experience little scattering from background gas. Second, UCPs may allow electron temperatures substantially lower than 10 K, and are therefore a potential way towards reaching the fundamental Fermi-degenerate limit (see Outlook).

\subsection{Initial design}\label{sec:initialdesign}
The UCES' seminal publication proposed using a ``dark" MOT \cite{ketterle1993high}, which allows atom densities up to \qty{e18}{m^{-3}} to extract a \qty{100}{pC} electron bunch from a \qty{1}{mm^3} volume, through near-threshold ionization with a nanosecond laser pulse. The narrow spectrum of the nanosecond laser pulse causes the electrons to have a very low initial momentum spread, associated with a bunch temperature on the order of \qty{1}{mK}. Next, electron disorder-induced heating (DIH) \cite{maxson2013fundamental} or the liberation of correlation energy \cite{kuzmin2002numerical} causes the temperature to rise to about \qty{10}{K} on the timescale of the plasma period or \qty{1}{ns}. 
The bunch is then to be extracted by an electric field of at least \qty{100}{MV/m}, which should be switched on in < \qty{1}{ns} using laser-triggered spark gap technology \cite{brussaard2004subnanosecond}. Particle tracking simulations show that combining this (rather extreme) set of parameters with uniformly filled ellipsoidal bunches, either `pancake' or `cigar' shaped, leads to a record peak brightness of $B_\perp = $ \qty{5e13}{A.rad^{-2}.m^{-2}} or \qty{5e14}{A.rad^{-2}.m^{-2}}, respectively.
The projected repetition rate would be 100 Hz, limited by the MOT loading rate.

\subsection{First version}
\begin{figure}
    \centering
    \includegraphics[width=0.8\linewidth]{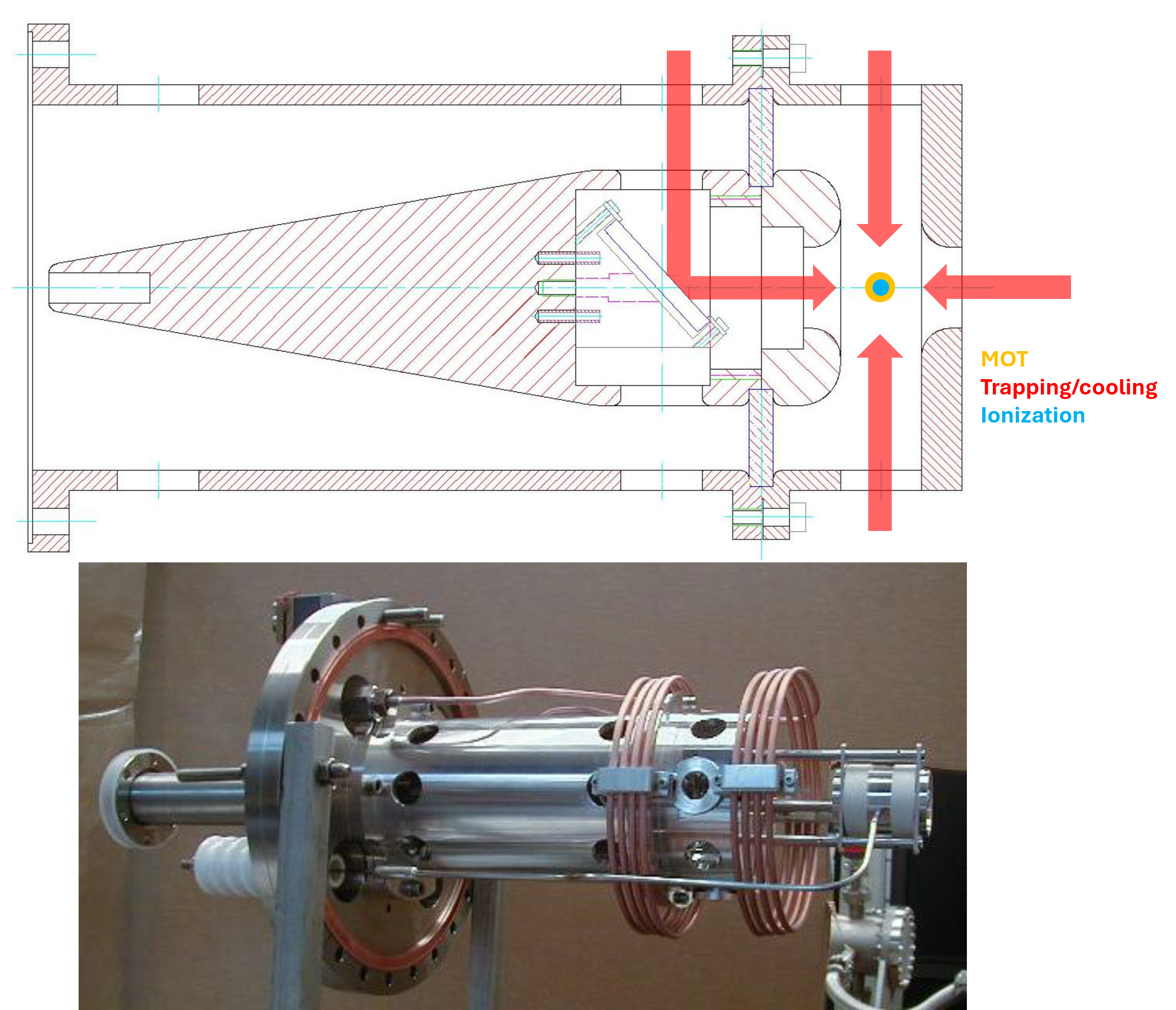}
    \caption{The first version of the UltraCold Electron Source. Top: schematic showing the cathode (tapered to reduce reflections in case of HV pulses with sub-ns rise times) and the MOT (orange) created by three pairs of counterpropagating laser beams (red) and a pair of coils in anti-Helmholtz configuration (not indicated). The ionization laser is indicated in blue. Bottom: Picture of the source, showing also the coils, and the MCP used to detect the electrons (right).}
    \label{fig:TabanFirstVersion}
\end{figure}

Needless to say, the first version of the source did not immediately reach these parameters. It did however produce record-breaking cold electron bunches. 
A relatively simple first try (\cite{claessens2007cold}) showed it was possible to extract either electrons ($\sim$ \qty{1.2}{pC} in \qty{200}{ns} rms) or ions ($\sim$ \qty{0.3}{pC} in \qty{800}{ns} rms) from a UCP. The plasma was created as follows: Rb-85 atoms were trapped and cooled in a MOT; The trapping beams were left on, thus populating the \textit{5p} state; a 480 nm, ns dye laser excited electrons to the \textit{44d} Rydberg state; The Rydberg atoms spontaneously ionize on a \unit{\us} timescale, forming a UCP (as described in \cite{robinson2000spontaneous}. An upper limit for the electron bunch temperature was estimated at \qty{500+-400}{K}. After this initial test, a first dedicated UCES was built.
As is shown in figure \ref{fig:TabanFirstVersion}, the entire source was built on a single vacuum flange. The MOT was created with three pairs of counterpropagating laser beams, in combination with a set of water-cooled, in-vacuum current coils in anti-Helmholtz configuration. In this first version, the trapping laser was kept on continuously, and no separate excitation pulse was used (skipping step \textbf{II} from the previous section). Electrons were thus created along a line, with a bunch charge of \qty{10}{fC}, using a \qty{2}{ns} (rms) laser pulse from a dye laser with an adjustable wavelength near the ionization limit $\lambda_0=$ \qty{479.06}{nm}, and accelerated by a constant potential applied to the cathode, of down to -30 kV. The electron bunch length was \qty{2}{ns} (rms), about equal to the laser pulse length.

\begin{figure}
    \centering
    \includegraphics[width=0.5\linewidth]{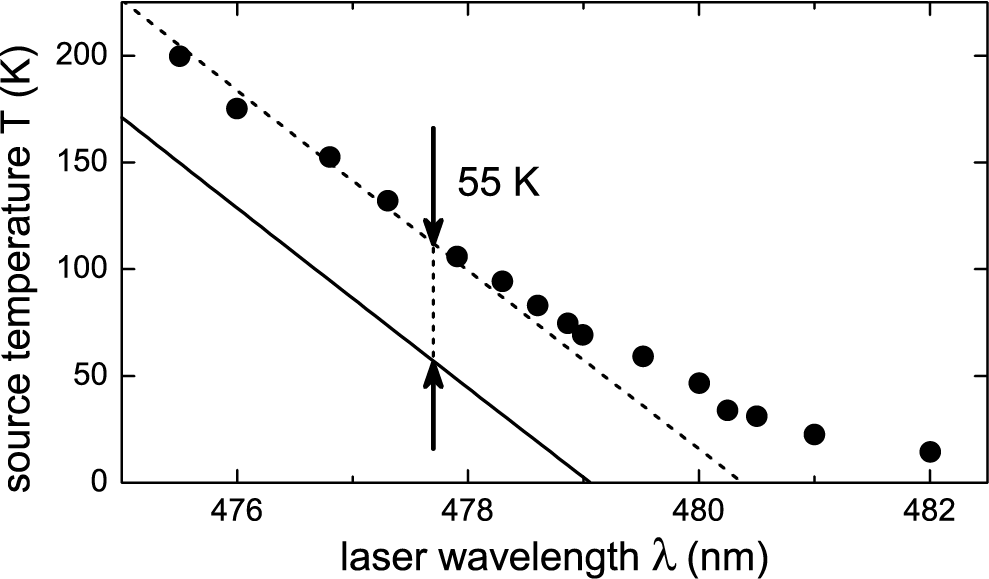}
    \caption{Source temperature as a function of ionization laser wavelength $\lambda$. The solid line represents the theoretical temperature due to the excess photon energy in absence of any electric field. The dots represent measured bunch temperatures and the \qty{55}{K} offset is (partially) attributed to the accelerating field, effectively reducing the ionization threshold. From \cite{Taban-thesis}.}
    \label{fig:TabanFirstVersionGraph}
\end{figure}

The bunch temperature was fully determined by the excess photon energy of the ionization laser, as is nicely shown in figure \ref{fig:TabanFirstVersionGraph}. This excess photon energy is given by 
\begin{equation}
    E_\lambda = hc \left ( \lambda^{-1} - \lambda_0^{-1} \right ),
\end{equation}
with $\lambda$ the wavelength of the ionization laser and $\lambda_0$ the field-free ionization limit given above, $h$ the Planck constant and $c$ the speed of light. As the ionization laser wavelength is increased, the excess photon energy is reduced. Down to about \qty{50}{K}, the slope equals $\frac{\mathrm{d}T}{\mathrm{d}\lambda} = \frac{2}{3}\frac{h c}{k_B}$ as expected assuming equal distribution of the excess energy over the 3 degrees of freedom. Below about \qty{100}{K} the temperature drops asymptotically, at the cost of reduced bunch charge, with the lowest measured bunch temperature being \qty{15}{K}. Note that the value for $\lambda_0$ stated above is for the field-free case, while in this experiment the applied accelerating field effectively reduces the ionization threshold, qualitatively causing the \qty{55}{K} offset indicated in the figure. This is known as a Stark shift, with the expression for the Stark shift energy given by
\begin{equation}
    E_F = 2 E_H \sqrt{\frac{F}{F_0}}, \quad \text{(Stark shift energy)}
\end{equation}
with $E_H=$\qty{27.2}{eV} the Hartree energy, $F$ the accelerating field strength and $F_0=$\qty{5.14e11}{V/m} the atomic unit of field strength. We can now write the total excess energy as:
\begin{align}\label{eq:E_exc}
    E_{exc} &= E_\lambda + E_F \nonumber\\
    &=hc \left ( \lambda^{-1} - \lambda_0^{-1} \right ) + 2 E_H \sqrt{\frac{F}{F_0}}.
\end{align}

In the same experiment, an alternative approach was also explored: here, the atoms were excited to Rydberg states and subsequently ionized by applying a pulsed field to the cathode (3 kV, 17 ns rise time), creating sub-ns electron bunches. This produced even colder electron bunches, with temperatures down to \qty{9}{K}.

In order to approach the brightness from the seminal paper, significant efforts were made to develop a pulsed HV source that would be able to switch \qty{1}{MV} with a sub-ns rise time, employing laser-triggered spark gap technology. Despite substantial progress \cite{taban2010ultracold} and previous experience with such switches \cite{Vyuga-thesis,Hendriks-thesis}, these goals proved technically too challenging, and the idea was abandoned. 

\subsection{Melbourne initiative}
\begin{figure}
    \centering
    \includegraphics[width=0.5\linewidth]{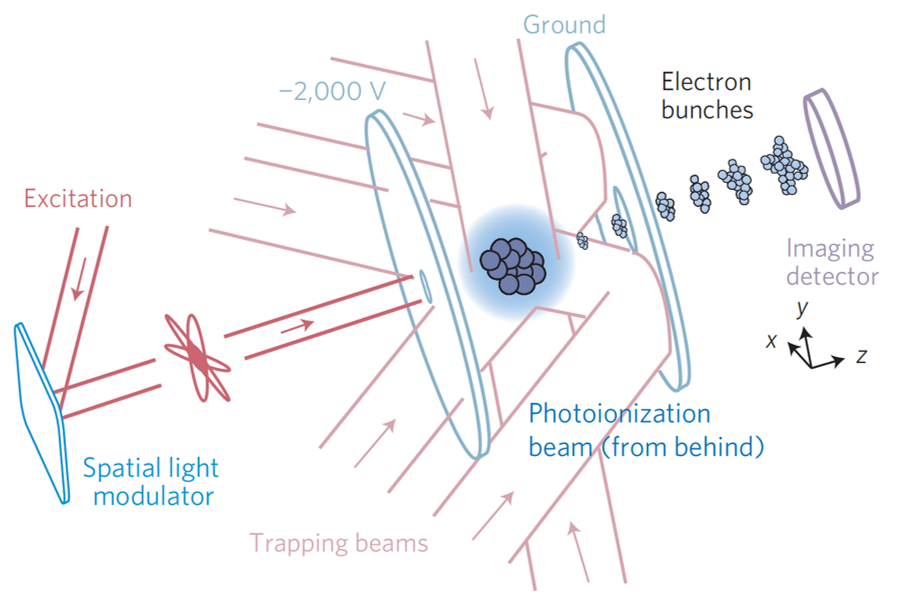}
    \caption{The setup used in Melbourne: Rb-85 atoms are trapped and cooled in a MOT in a DC accelerator, which accelerates the electrons produced through photoionization in the overlap volume between the ionization and excitation laser beam. The latter can be shaped with a spatial light modulator. From \cite{mcculloch2013high}.}
    \label{fig:McCulloch-setup}
\end{figure}

The exciting results described above proved contagious: the group of Rob Scholten at the University of Melbourne built their own ultracold electron source using a different geometry, shown in figure \ref{fig:McCulloch-setup}. Here, the Rb atoms are trapped and cooled in a MOT between two parallel plates which form the DC accelerator. The cathode is optically transparent, allowing the trapping laser beams to pass through, while the anode doubles as a mirror for the trapping laser beams (a design borrowed from \cite{knuffman2011nanoscale}). Excitation happens through a separate excitation beam, so ionization does not take place on a line, but in a volume defined by the overlap of an excitation laser (CW, switched on for about a \unit{\us}) and the ionization laser (5 ns), as explained in section \ref{sec:principle}. 
Their pioneering use of a Spatial Light Modulator (SLM) in the excitation beam enabled novel phase-space shaping experiments, as will be treated in section \ref{sec:phasespaceshaping}.
Consistent with the experiments in Eindhoven, bunch temperatures down to \qty{10+-5}{K} were reported, with a bunch length determined by the length of the ionization laser pulse, \qty{5}{ns}.

In order to bring this pulse length down and achieve ultrafast, cold electron bunches, the next step for both Eindhoven and Melbourne was to consider ionization by a femtosecond laser pulse.

\section{Youth: play and learn}\label{sec:youth}

\subsection{Femtosecond ionization}
There is a strong argument \textit{against} femtosecond ionization: when creating an ultracold plasma, the initial electron temperature (so on timescales shorter than the plasma period $T_p$) is determined by the spectral bandwidth of the ionization laser pulse, on the order of $E_{exc} = h\Delta f$.

For a Fourier-limited, Gaussian pulse, $(\Delta t)_{FWHM} (\Delta f)_{FWHM} = 2 \ln 2 /\pi$, so the electrons would then acquire a thermal distribution with $k_B T=\frac{\sqrt{\pi \ln2} \hbar}{\Delta t}$. For a 50 fs pulse this corresponds to \qty{225}{K}, compared to \qty{2}{mK} for the ns pulses used in the experiment above. This would mean that the curve traced out by the measured source temperature in figure \ref{fig:TabanFirstVersionGraph} would plateau at \qty{225}{K}, severely limiting the electron bunch coherence.

\begin{figure}
    \centering
    \includegraphics[width=0.6\linewidth]{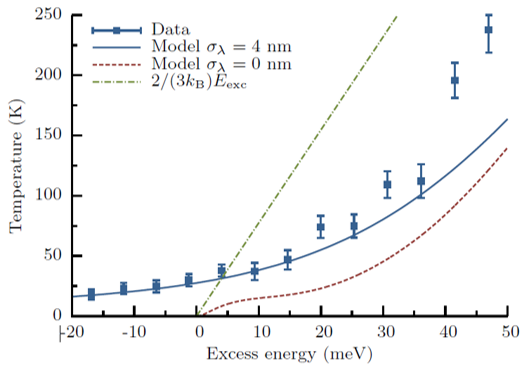}
    \caption{Femtosecond ionization: Experimentally measured bunch temperatures (blue squares) and the temperature expected from equipartition (green dash-dotted line). Also indicated are the curves for the analytical temperature model with and without taking the ionization laser spectral bandwidth into account (blue solid and red dashed resp.). From \cite{Engelen-thesis}.}
    \label{fig:Engelen-fs-temp}
\end{figure}

Luckily, the experiment showed that this picture is too simple.
In \cite{engelen2013high-coherence}, two important changes were made with respect to the first Eindhoven experiment. First, a \qty{50}{fs} ionization laser pulse was used, from an optical parametric amplifier (OPA) pumped by a Ti:Sa laser system. Second, ionization only took place in the overlap volume of the excitation laser and the ionization laser. The electron bunch temperature was then determined through a waist scan. The blue squares in figure \ref{fig:Engelen-fs-temp} show the obtained temperatures for different excess energies, with a minimum at \qty{18+-4}{K}. Later, these measurements were repeated in \cite{franssen2017pulse} with a higher-resolution detector and a higher accelerating field ($F=$ \qty{0.813}{MV/m}) leading to a minimum bunch temperature of \qty{10}{K}. These values are well below the limit expected from the spectral bandwidth of the ionization pulse (see \cite{engelen2014effective} for a detailed experimental comparison between fs and ns ionization).  
To understand how this is possible, we need to take a closer look at the electron emission process, as we will in the next section.
Around the same time, the Melbourne group combined femtosecond \textit{excitation} with nanosecond ionization \cite{mcculloch2013high} to produce short pulses. While they initially estimated the pulse length to be determined by energy spread at a minimum of \qty{150}{ps}, this was not supported by the measurements, and rectified by a later publication \cite{speirs2017identification}. Their ability to shape the excitation laser beam allowed formation of multiple beamlets, from which the emittance could be determined like in a pepper-pot measurement, for which a value of $\epsilon=$ \qty{538+-26}{nm} was found.

\subsection{Understanding the electron emission process}\label{sec:analyticalmodel}
\begin{figure}
    \centering
    \includegraphics[width=0.5\linewidth]{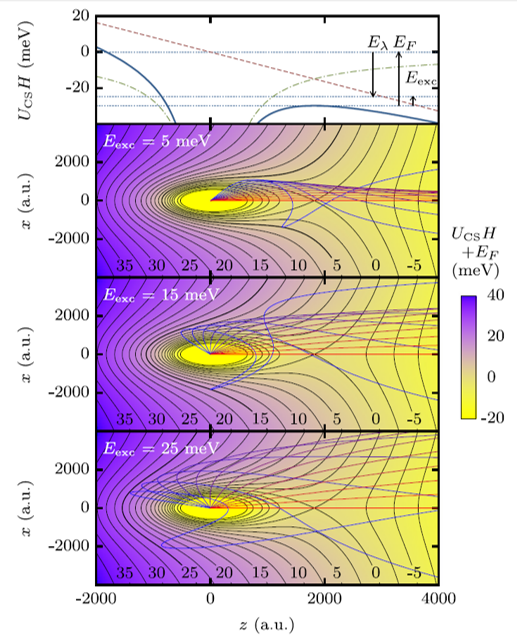}
    \caption{Modelling the near-threshold photo-emission from a single atom in an external electric field. Top: The field-free Coulomb potential (green dash-dot), the Stark potential (red dash) and the Combined Coulomb-Stark potential $U_{CS}$ (blue, solid), for $E_{exc}=$\qty{5}{meV} and $F=$\qty{-0.155}{MV/m}. Bottom: $U_{CS}$ potential landscape in $xz$-plane ($F$ in $z$-direction), with electron trajectories for $E_{exc}=$ 5, 15 and \qty{25}{meV} and starting angles $0\leq \beta < \beta_c$, where the color of the trajectory indicates $\beta$. From \cite{Engelen-thesis}.}
    \label{fig:Engelen-Ucs}
\end{figure}

The naive idea that the excess energy (Eq. \ref{eq:E_exc}) is distributed equally across all three degrees of freedom ignores the fact that the Stark shift is directional. For high photon energies this is irrelevant, but close to the ionization threshold $E_\lambda < E_F$ and so electron emission will be confined to a cone opening in the direction of the accelerating field. The Stark shifted atomic potential has a saddle point, with a minimum saddle point energy of $-E_F$. This is illustrated in the top panel of figure \ref{fig:Engelen-Ucs}, for an accelerating field $F$ pointing in the negative $z$-direction. From this top panel it is already clear that electron emission will be towards the positive $z$-direction. To gain quantitative understanding an analytical model is developed in \cite{engelen2014analytical}, based on a previously developed classical and non-relativistic model for ionization of a hydrogen atom \cite{kondratovich1984resonance, bordas1998classical}. 
With this model, closed analytical expressions are derived for the electron trajectories in a combined Coulomb-Stark potential $U_{CS}=-1/r+Fz$ (atomic units), as a function of $E_\lambda$, $F$ and $\beta$, where $\beta$ is the starting angle with respect to the $z$-axis. Only trajectories with a starting angle sufficiently close to the $z$-axis can escape the potential well, i.e. $\beta < \beta_c$, with $\beta_c = \cos^{-1}(-E_\lambda/E_F)$. In figure \ref{fig:Engelen-Ucs} these trajectories are drawn for different excess energies.
From these trajectories the (asymptotic) transverse velocity spread $\sigma_{v_x}$ and thus the transverse temperature can be calculated, assuming a probability distribution for $\beta$. Here, a uniform probability would correspond to an unpolarized laser pulse. For a linearly polarized pulse electrons are preferentially ejected along the laser polarization axis (\cite{engelen2013polarization}), following a $\cos^2\alpha$ distribution, with $\alpha$ the angle between the initial electron velocity and the laser polarization.
This assumes that the electron orbits of the state from which the atom is ionized (\textit{5p} in our case) do not have a preferential direction. This assumption is valid at the center of the MOT, where the magnetic field is zero and the magnetic quantization axis is therefore undefined.

The model uses a simple Coulomb potential, which is only correct for the hydrogen atom. For a more realistic Rb potential (\cite{engelen2014analytical}) the electron trajectories can no longer be described analytically and are calculated instead using the General Particle Tracer (GPT) code \cite{geer2011GPT}. In this potential no closed orbits exist, so all electrons (with $E_{exc}>0$) can escape the ion. Electrons with starting angle $\beta < \beta_c$ will escape the potential immediately as in the analytical model, while electrons with $\beta\geq\beta_c$ will make recursions in the potential until they scatter on the ion core. The electron will escape if the `new' $\beta$ after scattering is smaller than $\beta_c$. The resulting transverse temperature is therefore very close to that of the analytical model.

We can now use this model to quantitatively explain the transverse temperature from the femtosecond ionization experiment, as is drawn in red (dashed) in figure \ref{fig:Engelen-fs-temp}. An important correction is however still missing: the finite spectral bandwidth of the ionization laser pulse. This can be taken into account by convoluting the model curve with a Gaussian distribution corresponding to the spectral bandwidth, cut off at $E_{exc}=0$ (as no electrons are created for negative excess energies). This leads to the blue solid curve in figure \ref{fig:Engelen-fs-temp}.

Note that the analytical model is for a single-atom emitter, while in the experiments the transverse bunch temperature is measured from a collection of such emitters. For these measurements to be representative of the single-emitter dynamics, care was taken to avoid space charge effects by keeping the bunch charge low (typically a few hundred electrons).

\subsection{Longitudinal quality}\label{sec:longitudinalquality}
\begin{figure}
    \centering
    \includegraphics[width=0.55\linewidth]{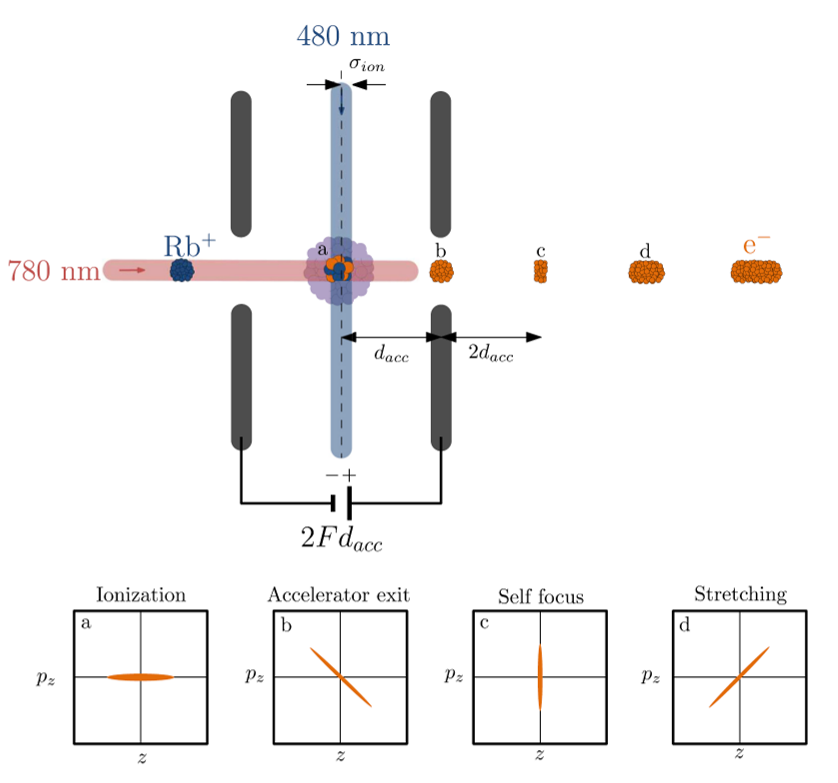}
    \caption{Schematic of the UCES, showing the excitation (780 nm) and ionization (480 nm) laser beams, and the electrons created in their overlap volume at a distance $d_{acc}$ from the ground electrode. The size of the overlap volume in the direction of the accelerator potential (determined by $\sigma_{ion}$) causes the electrons to self-compress at a distance $2d_{acc}$ behind the ground electrode. The diagrams below show the longitudinal phase space at different instances. From \cite{Franssen-thesis}.}
    \label{fig:franssen-selfcomp}
\end{figure}

Now that the effect of the emission dynamics on the transverse beam quality is understood, it is time to consider the longitudinal quality, i.e. the pulse length $\sigma_t$ and energy spread $\sigma_U$. First considering $\sigma_U$, two contributions can be distinguished: the (perfectly uncorrelated) energy spread due to the finite longitudinal temperature; and the (perfectly correlated) energy spread due to the length of the ionization volume in the direction of the accelerating field. From the analytical model above, we expect the longitudinal temperature to be much higher than the transverse one, but it will be limited by the total excess energy as
\begin{equation}\label{eq:nonequipart}
    E_{exc} = \frac{k}{2} \left ( T_x + T_y + T_z \right ),
\end{equation}
which for our cold electrons is typically 1-10 meV. The correlated energy spread due to the length of the ionization volume is given by
\begin{equation}
    \sigma_U = \sigma_{ion}eF,
\end{equation}
which is typically 0.1-1 eV, so at least an order of magnitude larger than the thermal contribution (this was also confirmed experimentally in \cite{franssen2018energy}).
As mentioned in Section \ref{sec:principle}, because of this correlated energy spread, the electron bunch of the UCES passes through a self-compression point at $2d_{acc}$ behind the DC accelerator. This is schematically depicted in figure \ref{fig:franssen-selfcomp}). At this longitudinal focus, $p_z$ and $z$ are uncorrelated, such that the pulse length consists of three contributions: the ionization laser pulse length $\tau_{laser}$, the spread in ionization delays $\tau_{ion}$, and a thermal contribution $\tau_{th}$ from the uncorrelated energy spread due to the finite longitudinal temperature. So
\begin{equation}\label{eq:tau_sc}
    \tau_{sc} = \sqrt{\tau_{laser}^2+\tau_{ion}^2+\tau_{th}^2}
\end{equation}
and the longitudinal emittance (equation \ref{eq:eps_z}) is then given by:
\begin{equation}
    \epsilon_z = \frac{\sigma_U}{mc}\sqrt{\tau_{laser}^2+\tau_{ion}^2+\tau_{th}^2}.
\end{equation}
The first, $\tau_{laser}$, is typically tens of fs (rms). The second, $\tau_{ion}^2$ can be estimated from the trajectories in the analytical model, and is typically a few picoseconds (depending on the $F$, $\lambda$ and the polarization of the ionization laser).
The third is given by
\begin{equation}
    \tau_{th} = \frac{\sqrt{mk_BT_z}}{eF}.
\end{equation}

In the first experimental characterization of the longitudinal quality of the electron beam, an RF cavity was used to measure the electron pulse length \cite{franssen2017pulse}. Due to geometric constraints, it was not possible to measure the pulse length in the self-compression point (but see Section \ref{sec:subps} for the follow-up measurement). Instead, the RF cavity had to be placed further downstream, such that the pulse length measurement was dominated by the energy spread $\sigma_U$.
Still, from the measured value of \qty{25}{ps} we can conclude that it is possible to produce electron bunches that are both ultracold and ultrafast.
In addition, a transverse temperature of \qty{10}{K} was measured, at an excess energy of $E_{exc}=$ \qty{10}{meV}, such that we can estimate $T_z$ using equation \ref{eq:nonequipart} to be about \qty{225}{K}, and $\tau_{th}=$ \qty{0.4}{ps}. 
We can thus safely conclude that the longitudinal emittance is dominated by the spread in ionization delays $\tau_{ion}$. Taking a value of $\tau_{ion}=$ \qty{2.5+-1}{ps} (as obtained from the analytical model for typical experimental values), the longitudinal emittance is $\epsilon_z=$ \qty{36+-14}{nm.rad} or $\epsilon_zmc=$ \qty{62+-24}{ps.eV}. 

\begin{figure}
    \centering
    \includegraphics[width=0.55\linewidth]{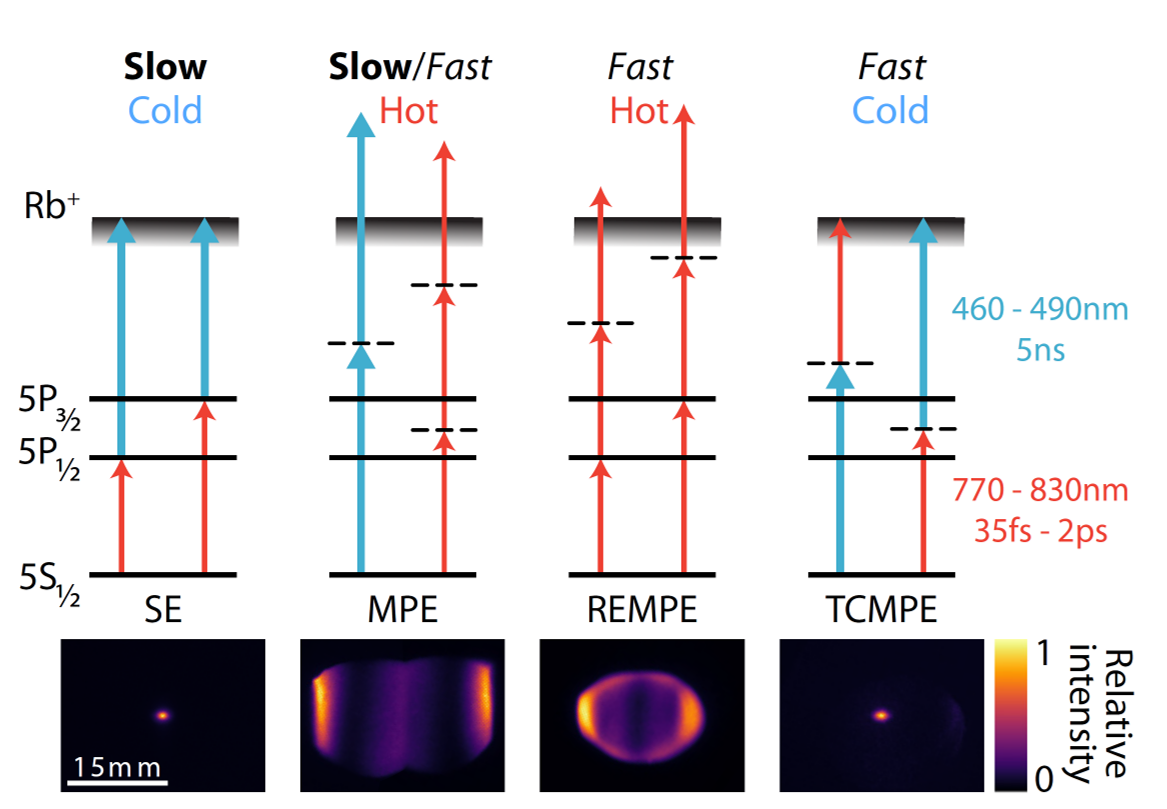}
    \caption{Top: The four ionization pathways that can occur in the Melbourne setup. Bottom: Measured transverse momentum distributions for the different pathways, showing that only SE and TCMPE generate cold bunches. From \cite{speirs2017identification}.}
    \label{fig:speirs-pathways}
\end{figure}

Around the same time, the Melbourne group also characterized their bunch lengths, using a pair of streaking electrodes. As the Melbourne setup employs fs excitation followed by ns ionisation, the possible ionization pathways differ from those in Eindhoven. This is reproduced in figure \ref{fig:speirs-pathways}. Four ionization pathways can be identified as follows. Sequential excitation (SE), as employed by the setup in Eindhoven, produces cold electrons but the pulse length is determined by the ionisation laser pulse length, as the lifetime of the \textit{5p} state is \qty{26}{ns}. With ns ionization, this thus produces ns electron bunches. Multiphoton excitation (MPE) and Resonance-enhanced multiphoton excitation (REMPE) occur at higher laser intensities, and produce hot electrons. Two-color multiphoton excitation (TCMPE) is the only possible route of generating electrons bunches that are both ultracold and ultrashort with this setup. Unfortunately, since this process is non-resonant, the ionization cross-section is three orders of magnitude lower than that of SE.

\subsection{Phase-space shaping}\label{sec:phasespaceshaping}
\begin{figure}
    \centering
    \includegraphics[width=0.45\linewidth]{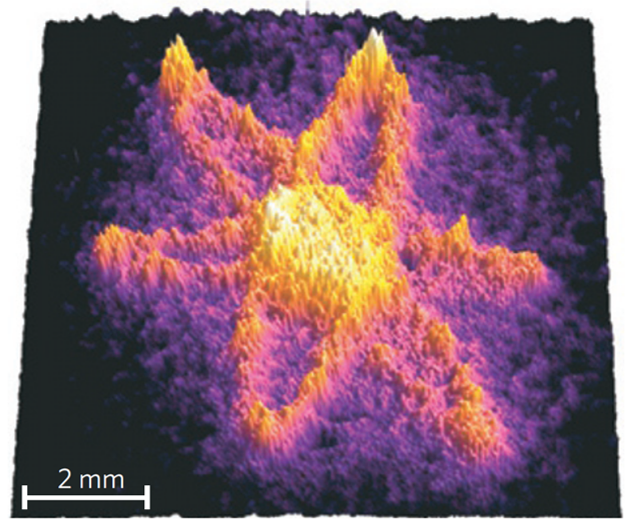}
    \caption{An electron bunch transversely shaped like an atom, measured ith the Melbourne setup. From \cite{mcculloch2011arbitrarily}.}
    \label{fig:mcculloch-atom}
\end{figure}

The fact that the electrons are created in the overlap of two perpendicular laser beams enables unique opportunities to control the initial shape of the electron bunch. The relatively simple `cigar' or `pancake'-shaped bunches - trading longitudinal for transverse coherence or vice versa - were already described in section \ref{sec:initialdesign}. But from the very start \cite{mcculloch2011arbitrarily}, the Melbourne group went much further: the incorporation of a spatial light modulator (SLM) in their excitation beam enabled generation of electron bunches with arbitrary transverse shapes. This is illustrated by the image in figure \ref{fig:mcculloch-atom}, which is the measured electron density of an electron bunch generated with an excitation beam in the shape of an atom. Thanks to the low bunch temperature, this shape is largely maintained during propagation down the beamline.
This is not just a fun trick, as the generation of initial bunches with sharp edges \cite{mcculloch2011arbitrarily} or consisting of multiple beamlets \cite{mcculloch2013high} allowed direct determination of the transverse emittance.

In a different experiment \cite{murphy2014detailed}, the Melbourne group used their shaped excitation beam to track space-charge dynamics. This was done on the ion bunches instead of the electrons, as this is much easier: their lower initial temperature (mK vs $\sim$ 10 K) and their higher mass make that the initial bunch structure is preserved for much longer, and that space charge dynamics are slowed down, approximately by a factor $\sqrt{m_{ion}/m_e}\approx$ 355. This means that the dynamics of their ion bunches of a few ns should be analogous to the dynamics of electron bunches of about \qty{10}{ps}. As shown in \cite{murphy2014detailed}, at the source nine closely spaced bunches were formed, and their Coulomb interactions give rise to compression layers between the bunches and phase-space wave-breaking at the boundary of the full beam, already at moderate bunch charges of a few fC.
Building on this result, and still employing ions, \cite{thompson2016suppression} compared bunch generation with different transverse profiles, showing how space-charge driven bunch expansion is self-similar for bunches with half-spherical or flat-top profiles, allowing a reduction of space-charge-driven emittance growth by up to 50 \% compared to a standard Gaussian profile.

In Eindhoven, efforts on shaping of the initial bunch distribution focused  rather on the longitudinal phase space: in \cite{franssen2017pulse} the ionisation laser was transversely shaped to create two foci, thus creating two electron bunches initially \qty{90}{\um} apart in the longitudinal direction. This was further developed in \cite{deRaadt-thesis}, by shaping the excitation beam using an SLM. These results open up the possibility of creating pulse trains, or microbunches, with applications in burst-mode UED or superradiant X-ray generation \cite{schaap2022photon}.

\subsection{The first diffraction experiments}
\begin{figure}
    \centering
    \includegraphics[width=0.8\linewidth]{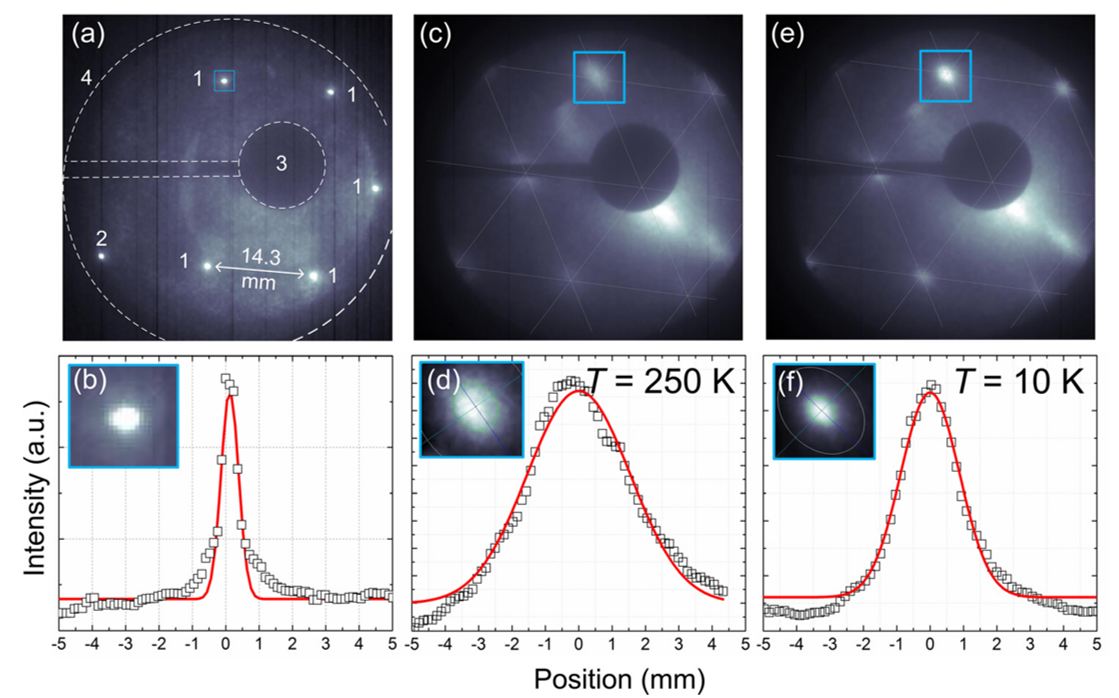}
    \caption{A first diffraction experiment with the UCES, on mono-crystalline graphite. (a): \qty{13.2}{keV} electrons are focused on the detector. Indicated are first order spots (1); a second order spot (2); a beam block (3); and the detector edge (4). (c): Diffraction pattern with the beam focused on the sample and a source temperature of \qty{250}{K}. (e): Idem for a source temperature of \qty{10}{K}. (b, d and f): One of the diffraction spots (inset) and a lineout along the minor axis of that spot, for each the patterns above. From \cite{vanMourik2014ultrafast}.}
    \label{fig:Mourik-Diff}
\end{figure}

With the peculiarities of the ultracold electron source now well understood, we can discuss the proof-of-principle diffraction experiments that both the Eindhoven and Melbourne groups performed around the same time. In (\cite{vanMourik2014ultrafast}), the ultracold electron beam was made to diffract of a graphite single crystal. Panel (a) in figure \ref{fig:Mourik-Diff} shows the first-order diffraction spots with the electron beam focused on the detector. In panel (c) and (e) the beam was focused to a tight, micron-sized spot on the sample instead, such that the size of the diffraction spots on the detector was dominated by the transverse coherence of the beam. By changing the ionization laser wavelength from 478 to 498 nm, the transverse temperature at the source is changed from \qty{250}{K} (c) to \qty{10}{K} (e), as measured using a waist scan (like in figures \ref{fig:TabanFirstVersionGraph} and \ref{fig:Engelen-fs-temp}). Together with a source size of $\sigma_{x,source}=$ \qty{32}{\um} these values correspond to a relative coherence of $C_{\perp,source}=$ \num{5.9e-5} and \num{2.9e-4} or, equivalently, an emittance of 6.6 and \qty{1.3}{nm.rad}. 
The increase in transverse coherence at the sample is evident from the increased brightness and reduced size of the diffraction spots.
From this diffraction spot size, the transverse coherence length can be calculated as 0.30 and \qty{0.55}{nm} resp., which together with the beam size $\sigma_{x,sample}$ (8.7 and \qty{3.3}{\um}) correspond to $C_{\perp,sample}=$ \num{3.4e-5} and \num{1.7e-4} or, equivalently, an emittance of 11 and \qty{2.3}{nm.rad}. Comparing the transverse coherence at the sample with that at the source (which are measured independently), we can see the effect of apparent emittance growth: a loss of coherence roughly on the order of a factor of two. This illustrates an important challenge in low-emittance beams: creating a low-emittance beam is only the first challenge, the second challenge is to preserve that beam quality down to the sample and detector. The two main causes for such emittance growth are space charge and imperfect electron optics. In these experiments, bunch charges were kept low (a few hundred electrons) to avoid space-charge effects, such that the emittance growth can be attributed to imperfect electron optics. Indeed, a simulation incorporating these imperfections (\cite{engelen2014effective}) reproduces the coherence at the sample to reasonable accuracy (Fig. 4 in \cite{vanMourik2014ultrafast}).

\begin{figure}
    \centering
    \includegraphics[width=0.7\linewidth]{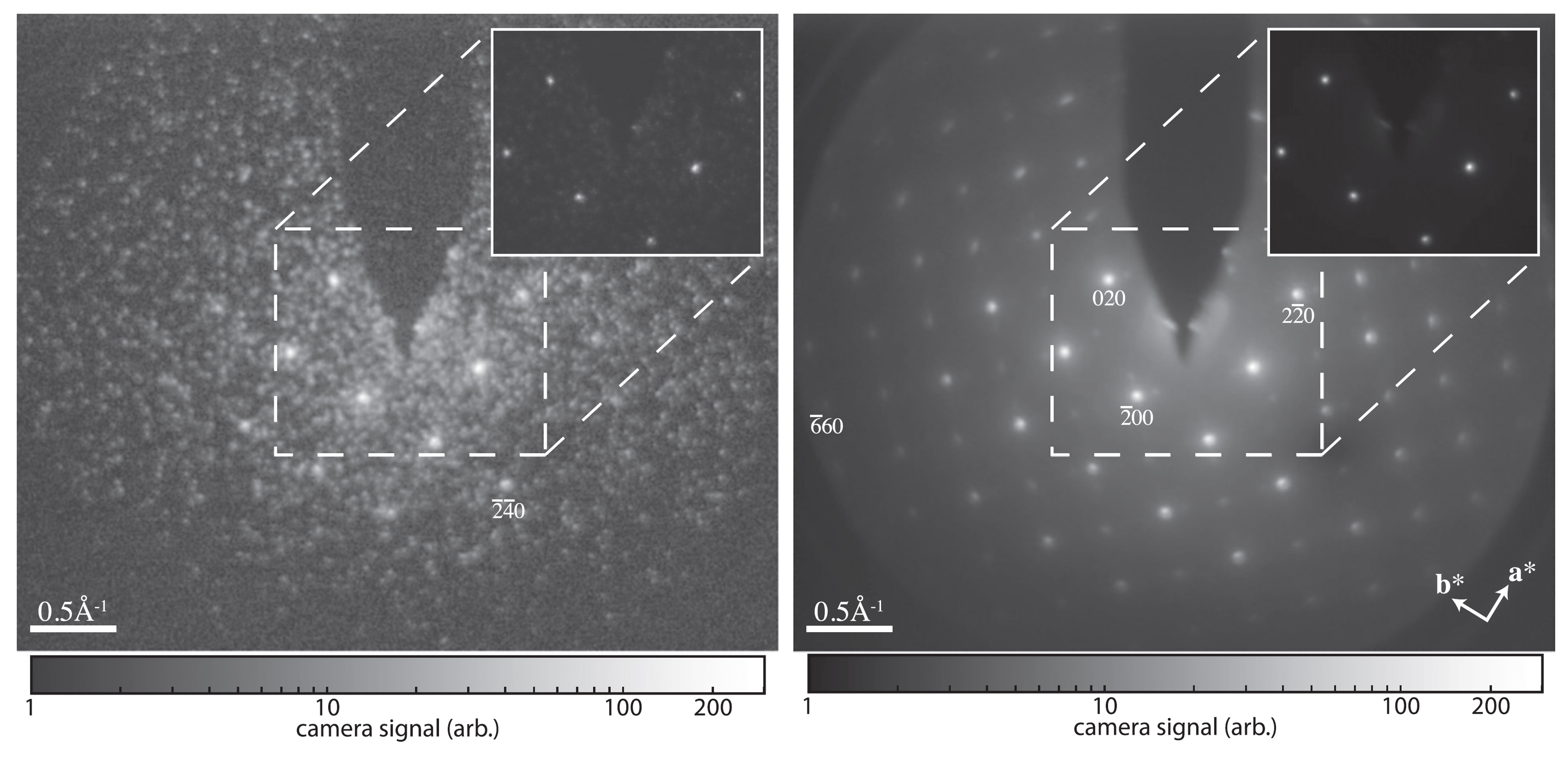}
    \caption{Left: Electron diffraction pattern of a gold crystal, from a single cold electron bunch. Right: Diffraction pattern obtained after registration and averaging of 2000 single-shot patterns. Grayscales of the main images are logarithmic, of the insets are linear. From \cite{speirs2015single}.}
    \label{fig:speirs-diff}
\end{figure}

Soon after, the Melbourne group also performed a diffraction experiment, with a significant improvement \cite{speirs2015single}. Their higher emittance and longer pulse length allow for a higher bunch charge before space-charge becomes problematic. Leveraging a bunch charge of \num{5e5} electrons, they performed single-shot electron diffraction of a gold crystal, as shown in the left panel of figure \ref{fig:speirs-diff}. These single-shot patterns contain enough signal to perform registration of successive patterns, thus allowing compensation for beam jitter and drift. A registered average of 2000 patterns is shown in the right panel of figure \ref{fig:speirs-diff}, with Bragg reflections up to ($\bar{6}60$).

\subsection{The Manchester AC-MOT}\label{sec:ACMOT}
We will end this section by mentioning the initiative in Manchester to build an ultracold electron source as well \cite{xia2014ultracold}. Uniquely, this source was to be based on the AC-MOT developed previously in Manchester \cite{harvey2008cold}. In an AC-MOT, an alternating current runs through the coils that generate the B-field required for atom trapping and cooling, such that the average B-field is zero. As a result, no net current is induced in the rest of the setup, allowing the B-field to be switched of in just tens of \unit{\us}, as opposed to several \unit{ms} for a standard DC MOT. Indeed, in both the Eindhoven and Melbourne setups, the B-field of the MOT is left on continuously, creating a focus (or even multiple foci) in the electron beam downstream. Additionally, keeping the B-field on also causes a hyperfine splitting of the energy levels of the cold atoms which increases with distance from the trap center. This splitting causes a (small) contributes to the spread in ionization energies, which can be eliminated by wwitching the B-field off before ionization. With the AC-MOT, the B-field can be switched off before each ionization pulse, and still operate at a 5-10 kHz repetition rate. According to the design the source was to generate 22 keV, 1 pC electron bunches at an emittance of \qty{0.35}{\um.rad} and to be the basis of an ultracold electron diffraction facility. This unfortunately never materialized. 

\section{Adolescence: Towards usable instruments}\label{sec:adolescence}
In this section we will see how the Eindhoven setup is transitioning from an experiment in itself towards a usable instrument. The activities in Melbourne unfortunately ceased after the results described in the previous section.

\subsection{Towards a compact, stable setup}
\begin{figure}
    \centering
    \includegraphics[width=0.75\linewidth]{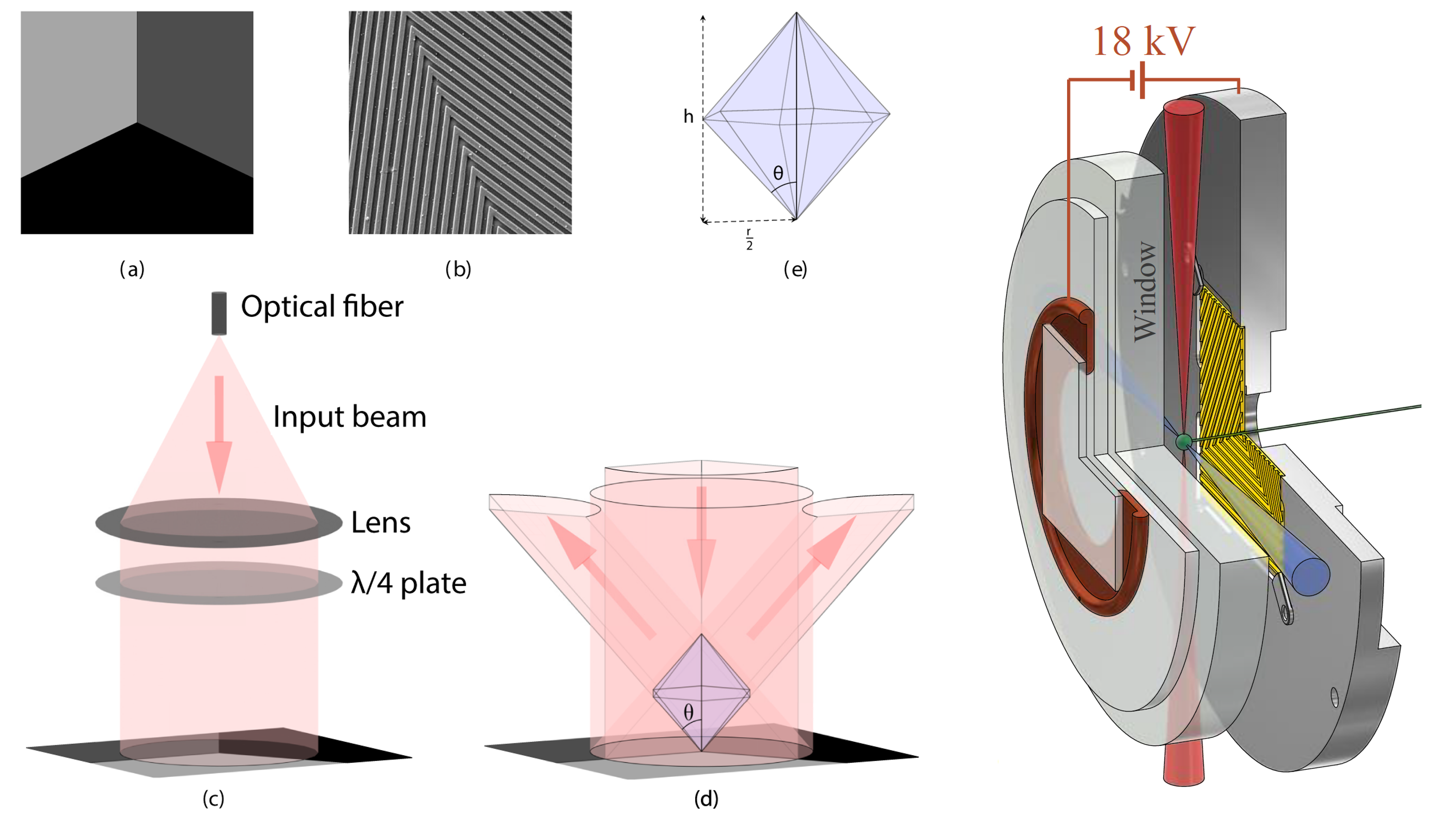}
    \caption{Left: Principle of the grating MOT. (a): Orientation of the three parts of the grating chip. (b): SEM image of part of the grating chip. (c): Simple setup of the grating MOT requiring only a single trapping beam. (d): Creation of the overlap volume between incoming and diffracted laser beams. (e): Geometry of the overlap volume. Right: The grating-MOT-based UCES. Adapted from \cite{Franssen-thesis} and \cite{deRaadt-thesis}.}
    \label{fig:gratingUCES}
\end{figure}

One of the main challenges in operating the setup of figure \ref{fig:TabanFirstVersion}, was the careful alignment of the six laser beams for trapping and cooling, in addition to the excitation and ionisation beams, in a configuration that also leaves room for the electron beam to exit the accelerator and ideally even for a camera to image the MOT. The next generation of the source considerable alleviated this challenge through the use of a grating MOT, a technique developed at the University of Strathclyde \cite{nshii2013surface}. This works as follows (see Fig. \ref{fig:gratingUCES}). A single trapping laser beam is used, which is incident upon a patterned chip (c). This patterned chip consists of three identical linear gratings, lying in a plane with \ang{120} relative angles (a and b). The incoming laser beam is diffracted by each grating, and these first-order diffracted beams partially overlap with the incoming beam to create a diamond-shaped overlap volume. The force balance between the incoming and diffracted beams can be used to trap and cool atoms \cite{mcgilligan2015phase}, similar to the conventional setup with 3 pairs of counterpropagating beams \footnote{The force balance is not identical: while for a conventional six-beam MOT the force balance is isotropic, for the grating MOT the axial force balance and radial force balance are typically different, except for a diffraction angle of $\arccos(1/3)$.}.

By embedding the grating chip in the ground electrode of a parallel-plate DC accelerator (right panel in figure \ref{fig:gratingUCES}), the principle of the grating MOT can be used to construct an UCES that is more compact and easier to align \cite{franssen2019compact}. It also allows easier optical access for the excitation and ionization beams as well as fluorescence imaging of the trapped atoms. This configuration additionally has the negative electrode of the DC accelerator outside vacuum, avoiding the need for a HV feedthrough. The electrons leave the accelerator through a laser-drilled hole in the center of the grating chip. The whole module is mounted on a single reentrant flange, while the MOT coils (not shown) are outside vacuum.
Commissioning of this `GMOT-UCES' is described in \cite{franssen2019compact}, which reports measurements of beam energies up to \qty{10}{keV} at an accelerating field of $F=$ \qty{0.37}{MV/m}. The normalized emittance was measured at \qty{1.9(0.9:1.5)}{nm.rad} which, with a source size of $\sigma_x=$ \qty{30}{\micro m}, corresponds to a transverse temperature of \qty{25+-25}{K}.
Although seemingly compact and convenient, this setup initially suffered from a major drawback: the RB$^+$ ions created at every shot impinge on the vacuum window of the accelerator. Because of the Rb background pressure, that window surface is covered with a thin layer of Rb atoms, which act as a cesiated cathode surface. As the accelerated Rb$^+$ ions hit this surface, a bunch of secondary, hot electrons is generated and accelerated down the beamline, often disturbing the measurement. Additionally, the (dielectric) window surface becomes gradually positively charged, thus gradually reducing the accelerator potential at the MOT position. This changes the electron beam energy, which in turn changes the beam's focal position and beam pointing. This was solved by implementing a fast HV-switch that quickly inverts the accelerator potential once the electron bunch has left the accelerator, slowing down the ions before they hit the vacuum window \cite{deRaadt-thesis}.
More recently, carefully designed mu-metal shielding was inserted around the electron beam, between the ground electrode and the sample plane. This greatly reduces the focusing effect of the MOT coils, avoiding the unwanted foci mentioned in section \ref{sec:ACMOT}.

\subsection{Ultracold and ultrashort: sub-picosecond UCES}\label{sec:subps}
\begin{figure}
    \centering
    \includegraphics[width=0.9\linewidth]{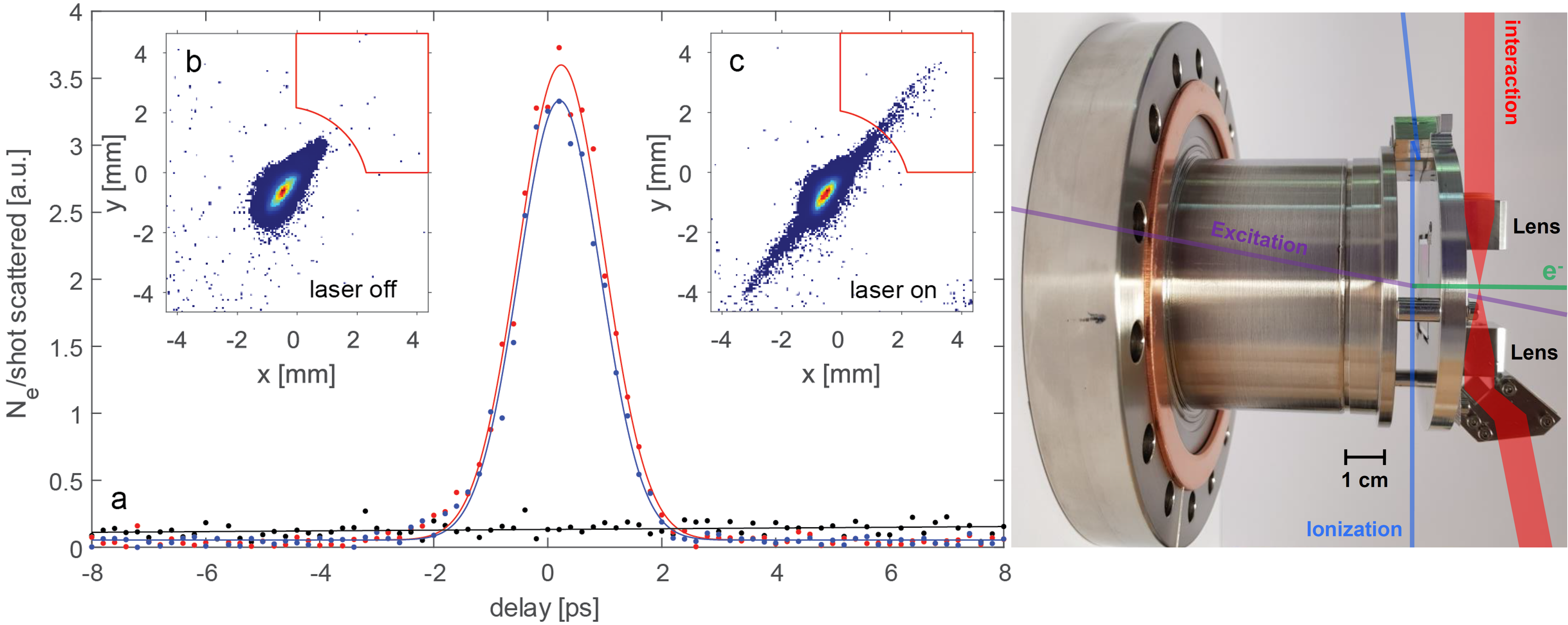}
    \caption{Left: The result of ponderomotive measurement of the pulse length at the self-compression point. In (a) the number of scattered electrons is plotted as a function of delay between the interaction laser and the electron bunch. The ionization laser was polarized either perpendicular (red) or parallel (blue) to the accelerating field. The background signal (no ionization laser) is plotted in black. The insets show typical electron beams on the detector with (c) and without (b) interaction laser. Electrons are counted as ponderomotively scattered if they are detected in the area outlined in red. Right: A picture of the reentrant flange with the DC accelerator. For the ponderomotive measurement two lenses are mounted on the back of the ground electrode, to focus the interaction laser (red). Adapted from \cite{deRaadt-thesis}.}
    \label{fig:ponderomotive}
\end{figure}

The improved ease of optical access offered by the new setup finally enabled a measurement that was long overdue: the pulse length at the self-compression point. 
In \cite{deRaadt2023subpicosecond} this pulse length was measured using a ponderomotive scattering technique, with a setup depicted in Fig. \ref{fig:ponderomotive} (right). By mounting two lenses on the back of the ground electrode of the DC accelerator (at a distance $2d_{acc}$ behind the front of the electrode) it was possible to focus a femtosecond laser pulse (`interaction' laser) down to a \unit{\micro m}-sized spot, reaching a peak intensity of about \qty{1.1e16}{W/cm^2}. When this interaction laser overlaps with the electron beam, part of the electrons are ponderomotively scattered in the transverse direction, which is visible on the electron detector downstream. The graph in Fig. \ref{fig:ponderomotive} (left) shows the number of scattered electrons as a function of delay between the interaction laser and the electron bunch. This yields an rms pulse length of $\tau_{sc}=$ \qty{735+-7}{fs}. The measurement was performed for polarizations parallel and perpendicular to the accelerating field (resp. blue and red\footnote{Parallel and perpendicular were erroneously swapped in \cite{deRaadt2023subpicosecond}.}), and for six different wavelengths of the ionization laser, i.e. for different excess energies.
Consistent with the electron emission model described above, the shortest $\tau_{sc}$ (Fig. \ref{fig:ponderomotive}) corresponds the case where the entire spectrum of the ionisation laser is above $hc/\lambda_0$, such that only electrons with an excess energy larger than the Stark shift are produced, which can directly leave the atom potential. Here, the departure from equipartition due to the directional Stark shift explained in the model of section \ref{sec:analyticalmodel} no longer applies, so these electrons are hot. Indeed, the central wavelength of the measurement of Fig. \ref{fig:ponderomotive} is \qty{474.4}{nm}, well below the Stark-shifted ionization threshold of 495 nm. This corresponds to an excess energy of \qty{111}{meV} or an (equipartitioned) temperature of \qty{857}{K} (while \cite{deRaadt2023subpicosecond} erroneously uses an equipartitioned temperature of 25 K in simulations and the calculation of $\epsilon_z$). 
As the spectrum is gradually shifted to lower photon energies, to where most electrons have negative excess energy, $\tau_{sc}$ broadens to \qty{1}{ps} with the addition of a few-ps tail \cite{deRaadt2023subpicosecond,deRaadt-thesis}. This confirms the estimates from Sec. \ref{sec:longitudinalquality} and \cite{franssen2017pulse}, that the longitudinal quality of the ultracold electron bunches is determined by the spread in ionization times, which is on a few-ps timescale. This means that there is room to minimize the electron excess energies by reducing the bandwidth of the ionization laser pulse by at least an order of magnitude, thus matching $\tau_{laser}$ to $\tau_{ion}$, at the cost of a minor increase in longitudinal emittance.

\subsection{Increasing the electron energy}\label{sec:DCRF}
Other than the stability and ease of use addressed by the new setup, another point requires attention in the transition from an experiment to a useful instrument: the electron energy. While the 1-10 keV energies reported above can be used for experiments in reflection (e.g. SEM or RHEED \cite{speirs2015single}), experiments in transmission geometry require much higher energies for all but the thinnest (few nm) samples. Unfortunately, the required presence of a MOT limits the accelerating field that can be applied, due to the induced Stark shift of the atomic transitions used for trapping and cooling. The GMOT UCES uses the highest accelerating field so far (\qty{1.2}{MV/m}).

In \cite{nijhof2023rf} an RF-cavity was used to post-accelerate the electrons produced by the GMOT-UCES to up to 35 keV, while the transverse quality of the electron bunches was derived from the diffraction pattern of a single gold crystal. The transverse emittances at the sample were found to be between 3-\qty{10}{nm.rad}, mainly attributed to imperfect beam dynamics in the RF-cavity, which was originally designed as a compression cavity for \qty{100}{keV} electrons \cite{vanoudheusden2010compression}.

\begin{figure}
    \centering
    \includegraphics[width=0.9\linewidth]{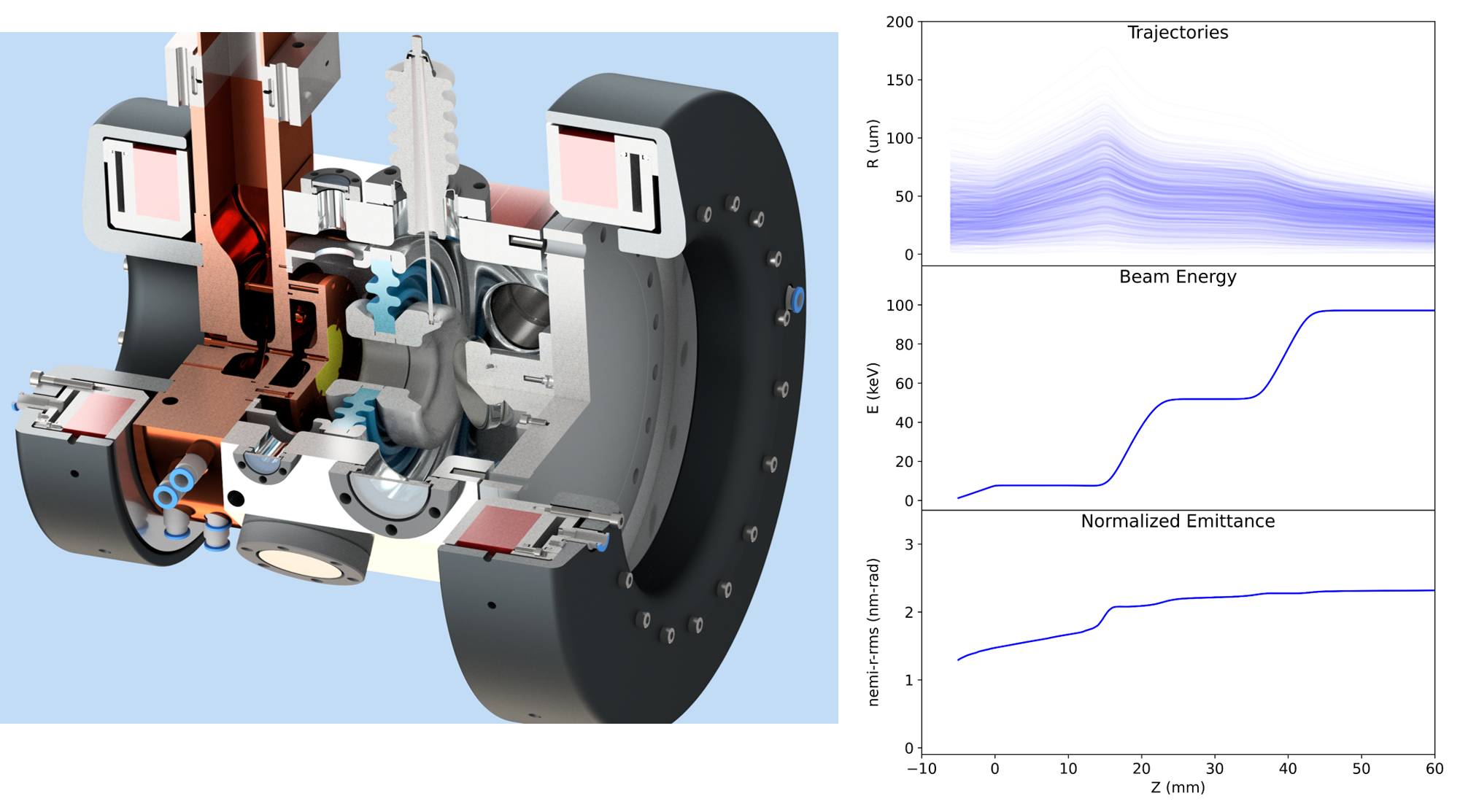}
    \caption{Left: CAD model of the ultracold DCRF source, illustrating the complexity of the setup. Right: Preliminary particle tracking simulations. (Work by A. Patwardhan, subject of a future publication by A. Patwardhan \textit{et al.})}
    \label{fig:DCRF}
\end{figure}

A more promising attempt is a dedicated setup that combines DC and RF acceleration. A complete design for such a setup was recently proposed (\cite{nijhof2023DCRF}). After further improvements, the ultracold DCRF source is currently in production. Figure \ref{fig:DCRF} (left) shows the final design, illustrating the complexity that results from the competition between optical, RF, vacuum, high-voltage, thermal and mechanical constraints.  A set of preliminary particle tracking simulations (using GPT) shows that a beam energy of 100 keV is reached with limited emittance growth.

\section{Conclusion: Lessons learned}\label{sec:conclusion}
In summary, the ultracold electron source has evolved into a stable, reliable setup, with a well-characterized electron beam. The fact that the electron beam is generated in a 3-dimensional ionization volume (rather than at a two-dimensional cathode surface) allows for a unique level of control over the initial phase space distribution of the electron beam. A relatively simple, classical model of the ionization process explains the observed low transverse temperatures. In order to fully exploit the potential offered by the ultracold electron source, several research directions are still waiting to be explored.

\section{Outlook}\label{sec:outlook}
The development that is nearest in the future is that of the DCRF accelerator, expected to bring the electron energy up to 100 keV. Near-to-medium term applications that are envisioned are for example the application to UED and even protein crystallography experiments, injection into a dielectric laser accelerator\cite{england2014dielectric}, or as injector for inverse Compton scattering \cite{vanelk2025xrayscompacttunablelinacbased}.
Highly relevant to such applications is the potential benefit offered by emittance compensation approaches \cite{vanoudheusden2010compression}. 
More exotic would be the generation of spin-polarized electron beams, or beating disorder-induced heating through the use of optical lattices, Rydberg crystals, or even through photodetachment from an anion Coulomb crystal \cite{comparat2023anions}.
Clearly, the story of the ultracold electron source has only just begun.

\begin{acknowledgments}
J.V. Huijts is supported by the European Union’s Horizon 2020 research and innovation programme under the Marie Skłodowska-Curie grant
agreement No. 101066850, and by a Branco Weiss Fellowship - Society in Science, administered by the ETH Zürich.
\end{acknowledgments}

\bibliography{bibliography}

\end{document}